\documentclass[preprint,aps,floats,showpacs]{revtex4}
\usepackage{graphicx}
\usepackage{epsfig}
\usepackage{pstricks}
\usepackage{pst-coil}
\usepackage{amsmath}

\def\beq{\begin{equation}}
\def\eeq{\end{equation}}
\def\beqa{\begin{eqnarray}}
\def\eeqa{\end{eqnarray}}
\newcommand{\lsim}{\mbox{\raisebox{-.6ex}{~$\stackrel{<}{\sim}$~}}}
\newcommand{\gsim}{\mbox{\raisebox{-.6ex}{~$\stackrel{>}{\sim}$~}}}

\def\cm{\,{\rm cm}}
\def\km{\,{\rm km}}

\def\s{\,{\rm s}}
\def\yr{\,{\rm yr}}

\def\sec{\,{\rm sec}}

\def\mpc{\,{\rm Mpc}}

\def\erg{\,{\rm erg}}
\def\ev{\,{\rm eV}}
\def\kev{\,{\rm keV}}
\def\mev{\,{\rm MeV}}
\def\gev{\,{\rm GeV}}

\def\pev{\,{\rm PeV}}

\def\msun{\,M_\odot}
\def\nubar{\bar{\nu}}
\def\nue{\nu_e}
\def\numu{\nu_\mu}

\def\numubar{\nubar_{\mu}}

\def\egamma{\epsilon_\gamma}
\def\egammab{\epsilon_{\gamma b}}
\def\egammaprime{\epsilon^{'}_\gamma}

\def\ep{\epsilon_p}

\def\epb{\epsilon_{pb}}

\def\enu{\epsilon_\nu}
\def\enuob{\epsilon_\nu^{\rm ob}}
\def\enub{\epsilon_{\nu b}}
\def\enustar{\epsilon_{\nu *}}
\def\xie{\xi_e}
\def\xip{\xi_p}
\def\xiB{\xi_B}
\def\grb{{\rm GRB}}
\def\rgrb{R_\grb}
\def\snIbc{{\rm SNIbc}}
\def\ccsn{{\rm CCSN}}
\def\rccsn{R_\ccsn}
\def\rsf{R_{\rm SF}}
\def\fgrbccsn{f_{\grb/\ccsn}}
\def\fgrbsnIbc{f_{\grb/\snIbc}}
\begin{document}
\title{\bf Upper Limit on the Cosmic Gamma-Ray Burst Rate from High Energy 
Diffuse Neutrino Background}
\author{Pijushpani Bhattacharjee}
\email{pijush.bhattacharjee@saha.ac.in}
\author{Sovan Chakraborty}
\email{sovan.chakraborty@saha.ac.in}
\author{Srirupa Das Gupta}
\email{srirupa.dasgupta@saha.ac.in}
\author{Kamales Kar}
\email{kamales.kar@saha.ac.in}
\affiliation{Theory Division, Saha Institute 
of Nuclear Physics, 1/AF Bidhannagar, 
Kolkata 700064. India}
\begin{abstract}
\noindent
We derive upper limits on the ratio 
$\fgrbccsn (z)\equiv \rgrb(z) /\rccsn(z)\, 
\equiv\fgrbccsn(0)(1+z)^\alpha$, the ratio of the rate, $\rgrb$, of 
long-duration Gamma Ray Bursts (GRBs) to the rate, $\rccsn$, of 
core-collapse supernovae (CCSNe) in the Universe ($z$ being 
the cosmological redshift and $\alpha\geq 0$), by using 
the upper limit on the diffuse 
TeV--PeV neutrino background given by the AMANDA-II experiment in 
the South Pole, under the assumption that GRBs are 
sources of TeV--PeV neutrinos produced from decay 
of charged pions produced in $p\gamma$ interaction of protons 
accelerated to ultrahigh energies at internal shocks within GRB 
jets. For the assumed ``concordance model'' of cosmic 
star formation rate, $\rsf$, with $\rccsn (z) \propto \rsf (z)$, 
our conservative upper limits are 
$\fgrbccsn(0)\leq 5.0\times10^{-3}$ for $\alpha=0$, and 
$\fgrbccsn(0)\leq 1.1\times10^{-3}$ for $\alpha=2$, for example. 
These limits are already comparable to (and, for $\alpha\geq 1$, 
already more restrictive than) the current upper limit on this ratio inferred from other astronomical considerations, thus providing a 
useful   
independent probe of and constraint on the CCSN-GRB connection. 
Non-detection of a diffuse TeV--PeV neutrino background by the 
up-coming IceCube detector in the South pole after three years of operation, 
for example, will bring down the upper limit on $\fgrbccsn (0)$ to below 
few 
$\times10^{-5}$ level, while a detection will confirm the 
hypothesis of proton acceleration to ultrahigh energies in GRBs and will 
potentially  
also yield the true rate of occurrence of these events in the Universe.  
\end{abstract}
\maketitle
\newpage
\section{Introduction
\label{sec:intro}}
Detection of supernova (SN) features in the 
afterglow spectra of several long duration 
(typically $>2\s$) Gamma Ray Bursts (GRBs) in the 
past one decade has provided strong support to the hypothesis that 
a significant fraction, if not all, of the long duration 
GRBs arise from collapse of massive stars; see, e.g., 
Refs.~\cite{Meszaros_araa02,Woosley-Bloom_araa06,DellaValle06} for 
recent reviews. The observed SN features in the 
GRB afterglow spectra are similar to those usually associated with 
core-collapse supernovae (CCSNe) of 
Type Ib/c (see, e.g., ~\cite{Woosley-Heger06,DellaValle06}). The total 
energy (corrected for beaming) in keV--MeV gamma rays emitted by typical 
long-duration GRBs is of order $10^{51}\erg$, which is roughly 
the same as the total explosion energy seen in typical CCSNe, 
although there exists considerable diversity in the energetics of 
both the SN and the GRB components in the SN-GRB associations 
observed so far. In particular, the estimated explosion energies of the 
SNe associated with the GRBs observed so far seem to be somewhat larger 
than those of normal SNe, leading to this ``special'' class of SNe 
being sometimes referred to as ``hypernovae''. 

The broad class of observational results on SN-GRB associations 
can be understood within the context of the ``collapsar'' 
model~\cite{collapsar_model_refs} in terms of a simple 
phenomenological picture (see, e.g., 
~\cite{Woosley-Zhang_astroph_0701320}) in which the 
core-collapse of a massive Wolf-Rayet star gives rise to two kinds 
of outflows emanating from the central regions inside the collapsed 
star: (a) a narrowly collimated and highly relativistic jet that is 
responsible for the GRB activity, the jet being driven, for 
example, by a rapidly rotating 
and accreting black hole formed at the center in the core-collapse 
process, and (b) a more wide-angled, quasi-spherical and 
non-relativistic (or at best sub-relativistic) outflow that goes to 
blow up the star and 
gives rise to the supernova. The energies channeled into 
these two components may in general vary independently, which may 
explain the diversity of energetics in the observed SN-GRB 
associations. Actually, depending on the energy contained in it the ``GRB 
jet'' may or may not be able to penetrate through the stellar material and 
emerge outside. Indeed, the fact that the 
SN-GRB associations observed so far involve CCSNe of Type Ib/c, 
but not of Type II, may be due to the inability of the GRB-causing 
jet to penetrate through the relatively larger amount of outer 
stellar material in the case of Type II SN as compared to that in 
SNe of Type Ib/c~\cite{fn1}. Considering various 
factors that may govern the energy channeled into the GRB-causing 
jet, such as the mass and rotation rate of the black hole, 
accretion efficiency, efficiency of conversion of accretion 
energy into collimated 
relativistic outflow, and so on, 
Woosley and Zhang~\cite{Woosley-Zhang_astroph_0701320} have 
obtained a rough lower limit of $\sim10^{48}\erg/\s$ for the power 
required for the jet to be able to emerge from the star. This is 
consistent with the energetics of the GRB components of the SN-GRB 
associations observed so far. 

While SN-GRB associations strongly support the stellar core-collapse 
origin of most long-duration GRBs, clearly, not all core-collapse events 
may result in a GRB --- 
the latter depends on whether or not the core-collapse event 
actually results in a 
``central engine'' (a rotating black hole fed by an accretion disk in the 
above mentioned phenomenological picture, for 
example) that is capable of driving the required collimated relativistic 
outflow. In other words, while every long-duration GRB would be expected 
to be accompanied by a 
core-collapse supernova~\cite{fn2}, the reverse is not true in general. 

What fraction of all stellar core-collapse events in the universe 
produce GRBs? Methods based on astronomical 
observations generally indicate the ratio between the cosmic 
GRB rate and the 
cosmic Type Ib/c SN rate, $\fgrbsnIbc$, to be in the range  
$\sim 10^{-3}$ -- $10^{-2}$ for a wide variety of different 
assumptions on various relevant parameters such as those that 
characterize the cosmic star 
formation rate (SFR), initial mass function (IMF) of stars, masses 
of Type Ib/c SN progenitors, the luminosity function of GRBs, 
the beaming factor of GRBs (associated with the fact that 
individual GRB emissions are 
highly non-isotropic and confined to narrowly collimated 
jets covering only a small fraction of the sky), and so on; see, 
for example, \cite{Guetta-DellaValle_06,Bissaldi_etal_07} and references 
therein. 
The dominant uncertainty in the estimate of $\fgrbsnIbc$ comes 
from the 
uncertainties in the estimates of the local GRB rate and the 
average GRB beaming factor. However, irrespective of the exact 
value of the ratio $\fgrbsnIbc$, 
it is clear that this ratio is significantly less than unity. This 
indicates that, apart from just being sufficiently massive stars, 
the GRB progenitors may need to satisfy additional special 
conditions. For example, it has 
been suggested~\cite{Woosley-Bloom_araa06} that the degree of 
rotation of 
the central iron core of the collapsing star and the metalicity 
of the 
progenitor star may play crucial roles in producing a GRB.  

In this paper, we discuss an alternative probe of the cosmic GRB rate 
that uses the predicted high energy (TeV--PeV) diffuse neutrino background 
produced 
by GRBs and the experimental upper limit on high energy diffuse neutrino 
background given by the AMANDA-II experiment in the South 
Pole~\cite{amanda_II_limit}. Existence of a high 
(TeV--PeV) energy diffuse GRB neutrino background (DGRBNuB) due to 
$p\gamma$ interactions of (ultra)high energy protons accelerated 
within GRB sources is a generic prediction~\cite{Waxman-Bahcall_grbnu} in 
most currently popular models of GRBs. This DGRBNuB 
is subject to being probed by 
the currently operating and up-coming large volume 
(kilometer scale) 
neutrino detectors such as IceCube~\cite{icecube}, 
ANITA~\cite{anita}, ANTARES~\cite{antares}, for example.  
Since neutrinos, unlike electromagnetic radiation, 
can travel un-hindered from the furthest cosmological distances, the 
DGRBNuB automatically includes the contributions from all GRBs in 
the Universe. Thus, an analysis of the DGRBNuB is likely to provide a 
good picture of the true rate of occurrence of these 
events in the Universe. Indeed, as we show in this paper, 
the upper limits on $\fgrbccsn (0)$, the ratio of the local 
(i.e., redshift $z=0$) GRB to CCSN rates, derived here from the 
consideration of DGRBNuB, are, for a wide 
range of values of the relevant parameters, already more
restrictive than the current upper limit on this ratio 
($\sim 2.5\times10^{-3}$) inferred from other astronomical 
considerations~\cite{Guetta-DellaValle_06,Bissaldi_etal_07}.  
Further, non-detection of a diffuse TeV--PeV neutrino background by the 
up-coming 
IceCube detector~\cite{icecube} in the South Pole after three 
years of operation, for example, will imply upper limits 
on $\fgrbccsn(0)$ at the level of few \ $\times10^{-5}$, while  
a detection of the DGRBNuB will 
provide strong support to the hypothesis of proton acceleration to 
ultrahigh energies within GRB jets. 

Our use of the DGRBNuB in constraining the cosmic GRB rate is in 
the same spirit as efforts to constrain the cosmic star formation rate 
(and thereby the cosmic CCSN rate) by using the experimental upper limit 
(set by the Super-Kamiokande (SK) detector)~\cite{SK_limit} on the 
predicted~\cite{dsnub_original} low (few MeV) energy Diffuse 
Supernova Neutrino Background (DSNuB); see, for 
example, Refs.~\cite{Ando-Sato_rev_04,Strigari_etal_05,Hopkins-Beacom_06}. 
Now that the cosmic SFR including its absolute normalization and 
thereby the cosmic CCSNe rate have got 
reasonably well determined by the 
recent high quality data from a variety of astronomical observations 
(see, e.g., ~\cite{Hopkins-Beacom_06}) 
(which, by the way, predicts a DSNuB flux that is  
close to the SK upper limit, implying 
that the DSNuB is probably close to being detected in the 
near future), one can begin to think of using this SFR to constrain 
the ratio of the cosmic GRB rate to CCSNe rate by using the predicted 
DGRBNuB flux together with the recent upper limits on the diffuse high 
energy neutrino flux from neutrino telescopes.    

We should emphasize here that the upper limits  
derived in this paper actually refer to the ratio of 
the rate of GRBs to that of {\it all} CCSNe including those of 
Type Ib/c and 
Type II, although SN-GRB associations observed so far involve 
SNe of Type Ib/c only. It is known, however, that Type II SNe 
probably constitute as much as $\sim$ 75\% of all CCSNe; 
see, e.g., \cite{Cappellaro_etal_07}. Thus, one can get the constraint on 
the GRB-to-SNIb/c ratio from the GRB-to-CCSNe ratio we obtain here by 
multiplying the latter by a factor of $\sim 4$. Conversely, for later 
comparison, we shall take the ``observed'' value of the ratio 
$\fgrbccsn(0)$ 
to be in the range $2.5\times (10^{-4}$ -- 
$10^{-3})$~\cite{Guetta-DellaValle_06,Bissaldi_etal_07}. 

Below, we first briefly review the calculation of the DGRBNuB 
spectrum 
in section \ref{sec:DGRBNuB_calc}. The resulting upper limits on 
$\fgrbccsn$ obtained by comparing the 
DGRBNuB with the current upper limit from AMANDA-II experiment are 
discussed in section \ref{sec:fgrbccsn_constraints} for various values of 
some 
of the 
relevant GRB parameters. Finally, in section \ref{sec:summary} 
we summarize the main results and conclude. 

\section{Diffuse high energy neutrinos from Gamma Ray Bursts
\label{sec:DGRBNuB_calc}}
Starting with the original calculations of Waxman and 
Bahcall~\cite{Waxman-Bahcall_grbnu}, the production of TeV--PeV neutrinos 
is widely accepted as a generic prediction of the fireball model of GRBs,  
provided, of course, that protons (in addition to electrons) are 
accelerated to ultrahigh energies within GRB jets. Reviews of the basic 
method of calculation of the expected neutrino flux from GRBs can be 
found, e.g., in ~\cite{Waxman_grb_rev_2001,Halzen-Hooper_rev_2002}.
Recent calculations of the GRB neutrino spectra can be found, for example,  
in ~\cite{Murase_etal_2006,Gupta-Zhang_dgrbnub}.   

For a given cosmological rate of occurrence of GRBs, the DGRBNuB flux can 
be calculated by simply convoluting the neutrino production spectrum of 
individual GRBs with the GRB rate density as a function of redshift, 
integrating over redshift up to some maximum redshift, and averaging over 
the intrinsic GRB parameters. In this paper we closely follow the recent 
calculation of the DGRBNuB spectrum described in 
Ref.~\cite{Gupta-Zhang_dgrbnub} with appropriate modifications for 
a possible enhanced evolution of the cosmic GRB rate in redshift 
relative to the cosmic SFR as indicated by a recent analysis of the 
{\it Swift} GRB data~\cite{Kistler_etal_07}. 

In the standard jet fireball model of GRBs (see, 
e.g., Refs.~\cite{Piran_grb_rev,Waxman_grb_rev_2001} for reviews), 
the fundamental source of the observed radiation from GRB is  
the dissipation of kinetic energy of 
ultra-relativistic (Lorentz factor $\Gamma \sim {\rm few}\,\, 100$) 
bulk flow of matter 
(caused by ejection from a ``central engine'')  
through formation of shocks which accelerate particles 
(electrons and 
protons) to ultra-relativistic energies. The shocks can form either 
inside 
the flow material itself due to collision of different shells of 
matter moving with different Lorentz factors (``internal shocks'')
 or due to collision of the flow material with an external medium 
(``external shocks''). The emission of the observed prompt 
$\gamma$-rays from a GRB source is attributed primarily to 
synchrotron radiation 
(with possible additional contribution from Inverse Compton 
scattering) of 
high energy electrons accelerated in the internal shocks.

It is expected that 
along with electrons, protons would also be accelerated at the 
internal 
shocks. Since synchrotron energy loss of protons is a slow process,  
protons can be accelerated to much higher energies than electrons. 
Indeed, it has been suggested~\cite{grb_uhecr_refs} that protons 
may be accelerated to ultra-high energies in GRB internal shocks 
and that these UHE protons may explain the observed ultra-high 
energy (UHE) cosmic rays (UHECR)~\cite{uhecr_obs} with energies 
up to $\sim10^{20}\ev$. 
These UHE protons interacting with the photons 
within the GRB jet would produce high 
energy charged pions through the photo-pion production process, $p+\gamma 
\to n+\pi^+$, and the subsequent decay of each charged pion would 
give rise to three high energy neutrinos (a $\numu$, a 
$\numubar$ and a $\nue$): $\pi^+\to \mu^+ + \numu\,,\,\, 
\mu^+\to e^+ +\numubar +\nue$. 

The dominant contribution to photo-pion production comes from the 
$\Delta$ resonance, $p+\gamma \to \Delta^+ \to n+\pi^+$,
 at which the $p\gamma$ interaction cross section peaks with a value 
$\sigma_{p-\gamma}^{\rm peak}\approx 5\times10^{-28}\cm^2$. 
This $\Delta$ resonance occurs at the proton threshold 
energy $\epsilon^{'}_{p,{\rm th}}$ (as measured in the GRB wind rest 
frame --- the ``comoving frame'' hereafter), which satisfies 
$\epsilon^{'}_{p,{\rm 
th}}\egammaprime\approx 0.3 \gev^2$, where 
$\egammaprime$ is the comoving frame energy of the colliding photon. 
In the rest frame of the GRB source (i.e., the central 
engine), the above threshold condition is 
$\epsilon_{p,{\rm th}}\egamma\approx 0.3\,
\Gamma^2\gev^2$, where $\Gamma$ is the bulk Lorentz factor of the 
GRB wind. In each $p\gamma$ interaction the 
pion takes away on average a fraction $\sim$ 20\% of the energy of 
the proton, so each neutrino from the decay of the pion carries 
$\sim$ 5\% of 
the energy of the initial proton, assuming that the four final state 
leptons share the energy of the decaying pion equally. Thus, for a 
typical photon energy $\egamma\sim1\mev$ and $\Gamma=300$, say, we have 
$\epsilon_{p,{\rm 
th}}\sim3\times10^7\gev$, which will give rise to neutrinos of 
energy $\sim1.5\pev$. 

The observed prompt $\gamma$ ray spectra of most GRBs are consistent
 with photon spectra which are well described by a broken 
power-law~\cite{Band_1993,Sakamoto_etal_2005}: $dn_\gamma/d\egamma 
\propto\egamma^{-\beta}$, with $\beta\approx 1.0$ for 
$\egamma < \egammab$, and  
$\beta\approx 2.25$ for $\egamma > \egammab$. For typical GRBs, 
the break 
energy $\egammab\sim 1\mev$. 
The normalized photon spectrum in the source rest frame (SRF) can be 
written as 

\beq
\frac{dn_{\gamma}}{d\egamma}
=0.2\, U_\gamma{\egammab}^{-1} \left\{ 
\begin{array}{l@{\quad 
\quad}l}{\egamma}^{-1} & {\rm for} \, \, \, 
\egamma\leq \egammab\,,\\
{\egammab}^{1.25}{\egamma}^{-2.25} & {\rm for} \, \, \, 
\egamma>\egammab\,,
\end{array}\right.
\label{photon_spect_1}
\eeq
where $U_\gamma$ is the total photon energy density in the SRF. 
Note that quantities in the SRF are related to those in the comoving 
frame (denoted by primes) by the appropriate powers of the Lorentz 
$\Gamma$ factor. Thus, for example, $\egamma=\Gamma \egammaprime$, and 
$U_\gamma=\Gamma^2\, U_\gamma^{'}=L_\gamma/4\pi r_d^2\, c$, where 
$L_\gamma$ is the photon luminosity in SRF, and $r_d=\Gamma^2 c 
t_v$ is the characteristic ``dissipation'' radius where internal shocks 
are formed and from where most of the radiation is emitted, 
$t_v\sim (10^{-2}$ -- $10^{-3}\, \sec)$ being 
the typical variability timescale of the emitted radiation. Note further 
that the quantities observed at earth (denoted by the superscript or 
subscript `ob') are related to those in the SRF 
through appropriate powers of the redshift factor $(1+z)$. Thus, for 
example, $\enuob=\enu/(1+z)$. 

We shall assume that at the internal shock protons are accelerated
to a differential spectrum, $dn_p/d\ep \propto \ep^{-2}$. 
The total internal energy in the system, $\mathcal{E}_{\rm total}$, is 
assumed to be distributed among electrons, protons and magnetic 
field as $\mathcal{E}_e=\xie \mathcal{E}_{\rm total}$, $\mathcal{E}_p=\xip 
\mathcal{E}_{\rm total}$ and 
$\mathcal{E}_B=\xiB \mathcal{E}_{\rm total}$, respectively, 
with $\xie + \xip + \xiB = 1$. We further 
assume that electrons are efficient radiators, so that 
$\mathcal{E}_e\approx \mathcal{E}_\gamma=\xie \mathcal{E}_{\rm total}$,  
where $\mathcal{E}_\gamma=L_\gamma \, T_d$ 
is the total isotropic-equivalent energy of the emitted gamma ray 
photons, $T_d$ being the total duration of the burst.    

It is worthwhile noting here that in the fireball model the kinetic energy 
of the initial bulk flow of matter is predominantly 
carried by protons, they being $\sim$ 2000 times more massive than 
electrons. This kinetic energy then is converted into internal energy 
at the shock, whereby the energy is now shared by protons, electrons and 
magnetic field. The mechanism by which the energy, which is initially 
carried mainly in the form of protons, gets 
transferred to electrons (and magnetic field) is not clear, but the 
phenomenology of the observed radiation from GRBs requires a significant 
fraction of the total internal energy to be eventually carried by 
electrons (see, e.g., ~\cite{Waxman_grb_rev_2001}). If this energy 
transfer from protons to electrons is very efficient, it may lead to 
equipartition of energy between them, i.e., $\xi_p=\xi_e$, but 
in general one may expect that $\xi_p/\xi_e \geq1$. 

Now, with the proton and photon spectra specified as above, 
the photo-pion production interactions of the protons with the 
photons given by the spectrum in eq.~(\ref{photon_spect_1}) can  
be shown to give rise to the neutrino 
spectrum~\cite{Waxman-Bahcall_grbnu,Gupta-Zhang_dgrbnub},
\beq
\enu^2\frac{dN_{\nu}(\enu)}{d\enu}\approx\frac{3}{8}\times 0.56\times 
f_{\pi}(\ep)\frac{\xip}{\xie}\mathcal{E}_{\gamma}
 \left\{\begin{array}{l@{\quad \quad}l} 1 & {\rm for} 
\,\,\, \enu<\enustar\,,\\
 (\enu/\enustar)^{-2} & {\rm for} \,\,\, \enu>\enustar\,,
\end{array}
 \right.
\label{nu_spect_1}
\eeq
where $f_\pi (\ep)$, the fractional energy loss of a proton to pions 
during the dynamical expansion time scale of the 
wind~\cite{Waxman-Bahcall_grbnu}, is to be evaluated at 
$\ep=20\enu$. For the photon spectrum given by 
equation (\ref{photon_spect_1}), $f_\pi (\ep)$ has the form 
~\cite{Gupta-Zhang_dgrbnub}      
\beq
f_{\pi}(\ep) =f_{0} \left\{\begin{array}{l@{\quad \quad}l}
0.88(\ep/\epb)^{1.25} & 
{\rm for} \,\,\, \ep < \epb\,,\\
1 & {\rm for}\,\,\, \ep > \epb\,,
\end{array} 
\right.
\label{f_pi_eqn}
\eeq
with $f_{0}=0.09 L_{\gamma,51}/(\Gamma_{300}^4\, t_{v,-3}\, 
\epsilon_{\gamma b,\mev})$. Here $L_{\gamma,51}=L_\gamma/(10^{51}~{\rm 
erg~s^{-1}})$, $t_{v,-3}=t_v/(10^{-3}~{\s})$, $\Gamma_{300}=\Gamma/300$, 
and $\epsilon_{\gamma b,\mev}=\egammab/\mev$.

In equation (\ref{nu_spect_1}) the 
factor $\frac{\xip}{\xie}\mathcal{E}_{\gamma}\approx
\xip \mathcal{E}_{\rm total}=\mathcal{E}_p$ is just the 
internal energy contained in 
protons, of which a fraction $f_\pi$ goes to pions. The factor 
$\frac{3}{8}$ comes from the 
fact that in $p\gamma$ interactions $\pi^+$'s and $\pi^0$'s 
are produced 
with roughly equal probability and the three neutrinos from the 
decay of $\pi^+$ together carry 3/4-th of the pion's energy. 
Finally, the factor 0.56 is an overall normalization factor. 

The spectrum (\ref{nu_spect_1}) has two breaks: The first break at 
$\enub=0.05\, \epb$ is caused by the break in $f_\pi (\ep)$ at 
$\epb$ with 
\beq
\epb=1.3\times10^{7}\, \Gamma_{300}^2\, 
(\epsilon_{\gamma b,\mev})^{-1}\, \gev\,, 
\label{epb_eqn}
\eeq
which, in turn, is due to the break in the photon spectrum 
(\ref{photon_spect_1}) at $\egammab$. 

The second break is at $\enustar$, with~\cite{Gupta-Zhang_dgrbnub}
\beq
\enustar=2.56\times10^{6}\xie^{1/2}\xiB^{-1/2}
L_{\gamma,51}^{-1/2}\, \Gamma_{300}^4\, t_{v,-3}\gev\,,
\label{enustar_eqn}
\eeq
which is due to muon cooling; for neutrino energy above $\enustar$ 
the corresponding energy of the parent muon (coming from the decay 
of the pion) would be high enough that the characteristic 
timescale of its 
energy loss through synchrotron radiation (``cooling") would be 
shorter than its decay time scale. Following 
\cite{Gupta-Zhang_dgrbnub} we shall assume $\xie=\xiB$, in which 
case $\enustar$ becomes independent of these two parameters. For a 
given $\mathcal{E}_\gamma$ (which is an observationally measurable 
quantity), the  
neutrino spectrum (\ref{nu_spect_1}) then depends on $\xip$ and 
$\xie$, but only through their ratio, $\xip/\xie$, which we shall 
take to be a free parameter in our calculations below.     

With the neutrino spectrum from individual GRBs (in the GRB 
source rest frame) given by equation (\ref{nu_spect_1}), the diffuse 
neutrino flux from all GRBs in the Universe, DGRBNuB, can be 
calculated as follows: 

Let $dn_\nu(\enuob)$ denote the present number density of neutrinos with 
energy between $\enuob$ and $\enuob + d \enuob$, which were emitted 
with energies between $\enu$ and $\enu+d\enu$ from 
GRBs at redshifts between $z$ and $z+dz$. Denoting by $\rgrb (z)$ the GRB 
rate per comoving volume at redshift $z$, we can write 
\beq
dn_\nu(\enuob)=\rgrb (z) (1+z)^3 \left(\frac{dt}{dz}dz\right)
\frac{dN_{\nu}(\enu)}{d\enu}\, d\enu\, (1+z)^{-3}\,. 
\label{dnuobs_eqn_1}
\eeq
In this equation the factor $(1+z)^3$ on the right hand side converts the 
GRB rate per comoving volume to the rate per physical volume while the 
factor $(1+z)^{-3}$ accounts for the dilution of the number density of 
the produced neutrinos due to expansion of the Universe.  

Using the standard Friedmann relation 
\beq
\frac{dt}{dz}=-\left[H_0(1+z)\sqrt{\Omega_m(1+z)^3 
+ \Omega_\Lambda}\right]^{-1}
\label{t-z_relation}
\eeq
(we shall use the standard $\Lambda$CDM cosmology parameters, 
$\Omega_m=0.3\,$, $\Omega_\Lambda=0.7\,$ and $H_0=70\km/\s/\mpc$), 
the total differential flux of neutrinos, $\Phi(\enuob)$, giving the 
number of neutrinos (of all flavors) crossing per unit area per unit time 
per unit energy per unit solid angle, due to all 
GRBs in the Universe up to a maximum redshift $z_{\rm max}$ 
can be written as 
\beqa
\Phi(\enuob) & \equiv & 
\frac{c}{4\pi}\frac{dn_\nu({\enuob})}{d\enuob}\nonumber\\
 & = & \frac{c}{4\pi}H_0^{-1}\int_0^{z_{\rm max}} \rgrb (z) 
\frac{dN_{\nu}(\enu)}{d\enu}\frac{dz}{\sqrt{\Omega_m(1+z)^3
+ \Omega_\Lambda}}\,,
\label{flux_eqn_1}
\eeqa
where $\frac{dN_{\nu}(\enu)}{d\enu}$ is given by equation 
(\ref{nu_spect_1}) with $\enu=(1+z)\enuob$. We assume  
that, because of the long cosmological baseline, neutrino flavor 
oscillation distributes the original neutrinos equally into 
all three flavors.  

The source spectrum $\frac{dN_{\nu}(\enu)}{d\enu}$ for a single GRB 
is a function of various GRB parameters: $L_\gamma\,,\, \Gamma\,,\,  
T_d\,,\, t_v\,,\, \xip/\xie\,$ and $\egammab$. We average over 
the ``measurable'' GRB parameters $L_\gamma\,,\, \Gamma\,,\,
T_d\,$ and $t_v$ using the procedure described in 
Ref.~\cite{Gupta-Zhang_dgrbnub} using the same distribution 
functions for these parameters used there~\cite{fn3}. 
The break energy $\egammab$ can be 
related to total energy in photons, $\mathcal{E}_\gamma$ 
(or equivalently to luminosity $L_\gamma$) through the empirical 
``Amati relation"~\cite{amati_rel} given by 
($\egammab/100\kev)=(3.64\pm0.04)(\mathcal{E}_\gamma/
7.9\times10^{52}\erg)^{0.51\pm0.01}$. And, as already mentioned, 
the ratio $\xip/\xie$ remains as a free parameter.  

What remains to be specified is the GRB rate as a function of 
redshift, $\rgrb (z)$. Stellar core-collapse origin of GRBs as 
evidenced by CCSN-GRB associations implies that GRB rate should 
follow CCSN rate, $\rccsn (z)$, which is proportional to SFR, 
$\rsf (z)$. Recently, however, an analysis~\cite{Kistler_etal_07}
of a reasonably large sample of GRBs with known redshifts from 
the {\it Swift} mission~\cite{swift_mission}, together with recent 
accurate determination of the star formation 
history~\cite{Hopkins-Beacom_06}, has given strong 
indication of a possible enhanced evolution of the GRB rate 
(with redshift) relative to SFR. Taking cue from this we shall 
allow for a possible effective evolutionary factor in the GRB rate 
relative to SFR and write 
\beq
\rgrb (z)\propto (1+z)^\alpha \rsf (z)\,,
\label{rgrb_rsf_rel_1}
\eeq
where $\alpha\geq0$ is a constant, and $\rsf (z)$ (rate per {\it comoving 
volume}) is taken as 
~\cite{Hopkins-Beacom_06,Kistler_etal_07}
\beq
\rsf (z) \propto\, \left\{
\begin{array}{lcl}
(1 + z)^{3.44} & {\rm for} & z < 0.97 \\
(1 + z)^{-0.26} & {\rm for} & 0.97 < z < 4.48 \\
(1 + z)^{-7.8} & {\rm for} & 4.48 < z \,,
\end{array}
\right.
\label{rsf_Hopkins-Beacom}
\eeq
with $\rsf (0)=0.0197\,M_\odot$~yr$^{-1}$~Mpc$^{-3}$. 
This SFR including its normalization has been derived from and is in 
concordance with recent accurate 
data on a variety of different indicators of SFR in the Universe, and is 
also in conformity with the experimental upper limit on the DSNuB flux 
given by the Super-Kamiokande experiment~\cite{SK_limit}. Following the 
terminology introduced in \cite{Strigari_etal_05} we shall refer to the 
above SFR as the ``concordance model'' of SFR. 

The core-collapse supernova rate, $\rccsn (z)$, is related to $\rsf (z)$ 
through the Initial Mass Function (IMF), $dn/dm$, giving the differential 
mass distribution of stars at formation. Thus,
\beq
\rccsn (z)=\frac{\int_{8\msun}^{100\msun}\frac{dn}{dm}\,
dm}{\int_{0.1\msun}^{100\msun}m\, \frac{dn}{dm}\, dm}\, \rsf (z)\,,
\label{rccsn_rsf_imf_rel}
\eeq
where, following standard practice, the IMF is
assumed to be epoch (redshift) independent (see, e.g.,
\cite{Ando-Sato_rev_04,Strigari_etal_05}), and we have assumed that all 
stars more massive than $\sim 8\msun$ undergo core-collapse and die on 
a time scale short compared to Hubble time. Also, our results 
are insensitive to the exact value of the upper cut-off of the IMF
(chosen to be at $100\msun$ above) as long as it is sufficiently
large ($\gsim 30\msun$ or so). The SFR (\ref{rsf_Hopkins-Beacom}) 
assumes an IMF of the form ~\cite{Baldry-Glazebrook}, $dn/dm \propto 
m^{-2.15}$ for $m > 0.5\msun$, and $dn/dm\propto m^{-1.50}$ for 
$0.1\msun \lsim m \leq 0.5\msun$. With this, the GRB rate can be written 
in terms of CCSN rate as 
\beq
\rgrb (z)\equiv\fgrbccsn(z)\rccsn(z)= \fgrbccsn (0)\,\, (1+z)^\alpha\,\, 
\rccsn (z)\,,
\label{rgrb_rccsn_rel}
\eeq
where the the normalized core-collapse 
event rate in the Universe, $\rccsn (z)$, using equations 
(\ref{rsf_Hopkins-Beacom}) and (\ref{rccsn_rsf_imf_rel}), is  
\beq
\rccsn (z) = 2.60\times 10^{-4}\yr^{-1}\mpc^{-3}\, \left\{
\begin{array}{lcl}
(1 + z)^{3.44}  & {\rm for} & z < 0.97 \\
12.29 \, (1 + z)^{-0.26} & {\rm for} & 0.97 < z < 4.48 \\
4.57\times10^6 \, (1 + z)^{-7.8} & {\rm for} & 4.48 < z \,. 
\end{array}
\right.
\label{rccsn_Hopkins-Beacom}
\eeq

The analysis of 
Ref.~\cite{Kistler_etal_07} seems to indicate the best-fit value 
of the evolution index $\alpha$ appearing in equations 
(\ref{rgrb_rsf_rel_1}) and (\ref{rgrb_rccsn_rel}) to be  
$\sim 1.5$, but in this paper we shall keep $\alpha$ as a 
free parameter and study the dependence of our derived upper limits on 
$\fgrbccsn (0)$ on $\alpha$. 

\section{Upper limits on $\fgrbccsn$
\label{sec:fgrbccsn_constraints}}
\begin{figure}[h]
\epsfig{file=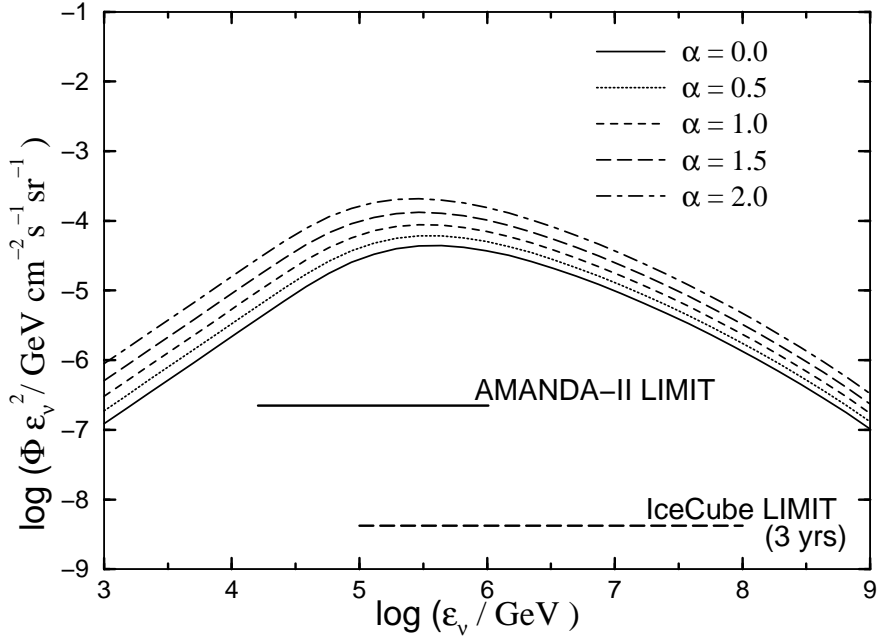,angle=270,width=4.5in}
\caption{The total (all flavor) DGRBNuB flux with 
$\fgrbccsn(0)=1$ for the equipartition case 
of $\xi_p/\xi_e=1$ for various values of $\alpha$, the effective evolution 
index of the cosmic GRB rate relative to cosmic star formation rate. The 
current 90\% C.~L.~upper limit on the diffuse neutrino flux given by the 
AMANDA-II experiment~\cite{amanda_II_limit} and the projected
upper limit from the IceCube experiment after three years of operation 
~\cite{icecube_3year} are also shown.}  
\label{fig:flux_equipart0.33}
\end{figure}

\begin{figure}[h]
\epsfig{file=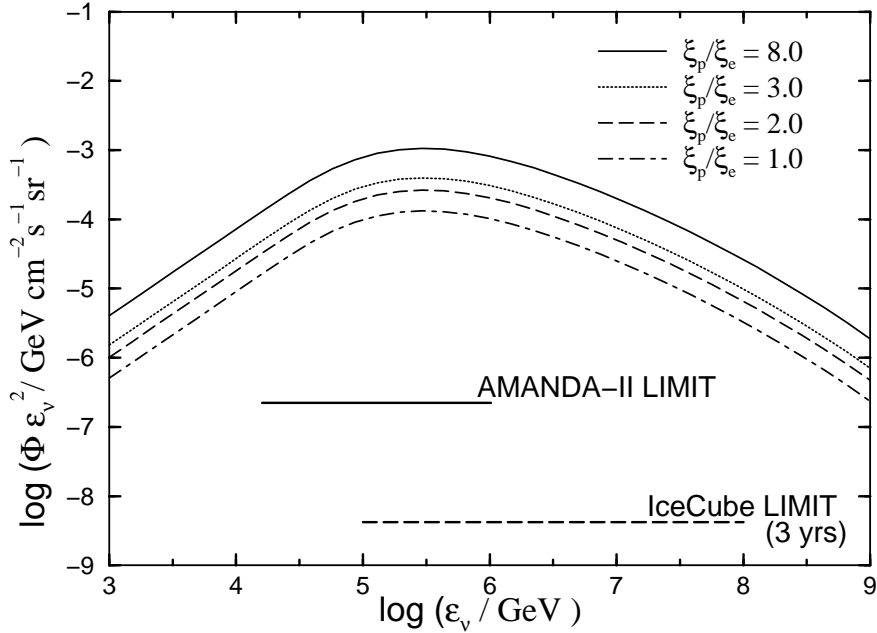,angle=270,width=4.5in}
\caption{Same as Fig.~\ref{fig:flux_equipart0.33}, but for the fixed 
value of $\alpha=1.5$, and various different values of the parameter 
$\xi_p/\xi_e$.} 
\label{fig:flux_alpha1.5}
\end{figure}

\begin{figure}[h]
\epsfig{file=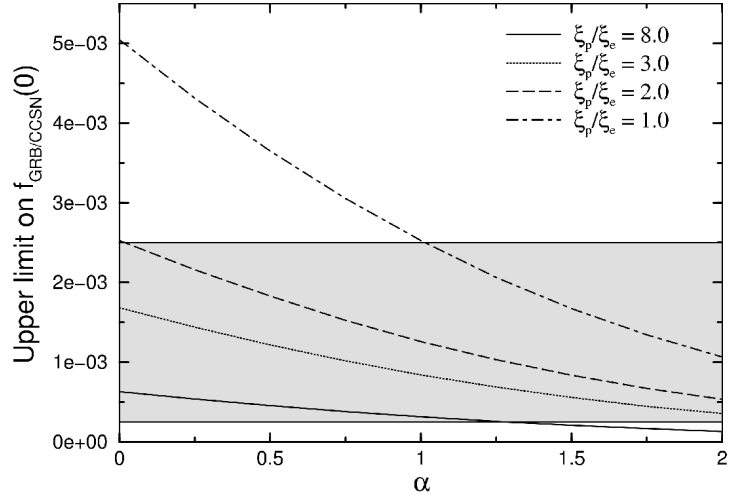,angle=270,width=4.5in}
\caption{Upper limits on the fraction $\fgrbccsn (0)$, obtained by 
requiring $\enu^2\Phi(\enu)$ for the DGRBNuB to not exceed the AMANDA-II 
limit, are shown as function of the GRB evolution index $\alpha$ for 
various values of the parameter $\xi_p/\xi_e$ . The shaded region 
indicates the range of values of $\fgrbccsn(0)$ estimated from 
other astronomical considerations. } 
\label{fig:f_vs_alpha}
\end{figure}

\begin{figure}[h]
\epsfig{file=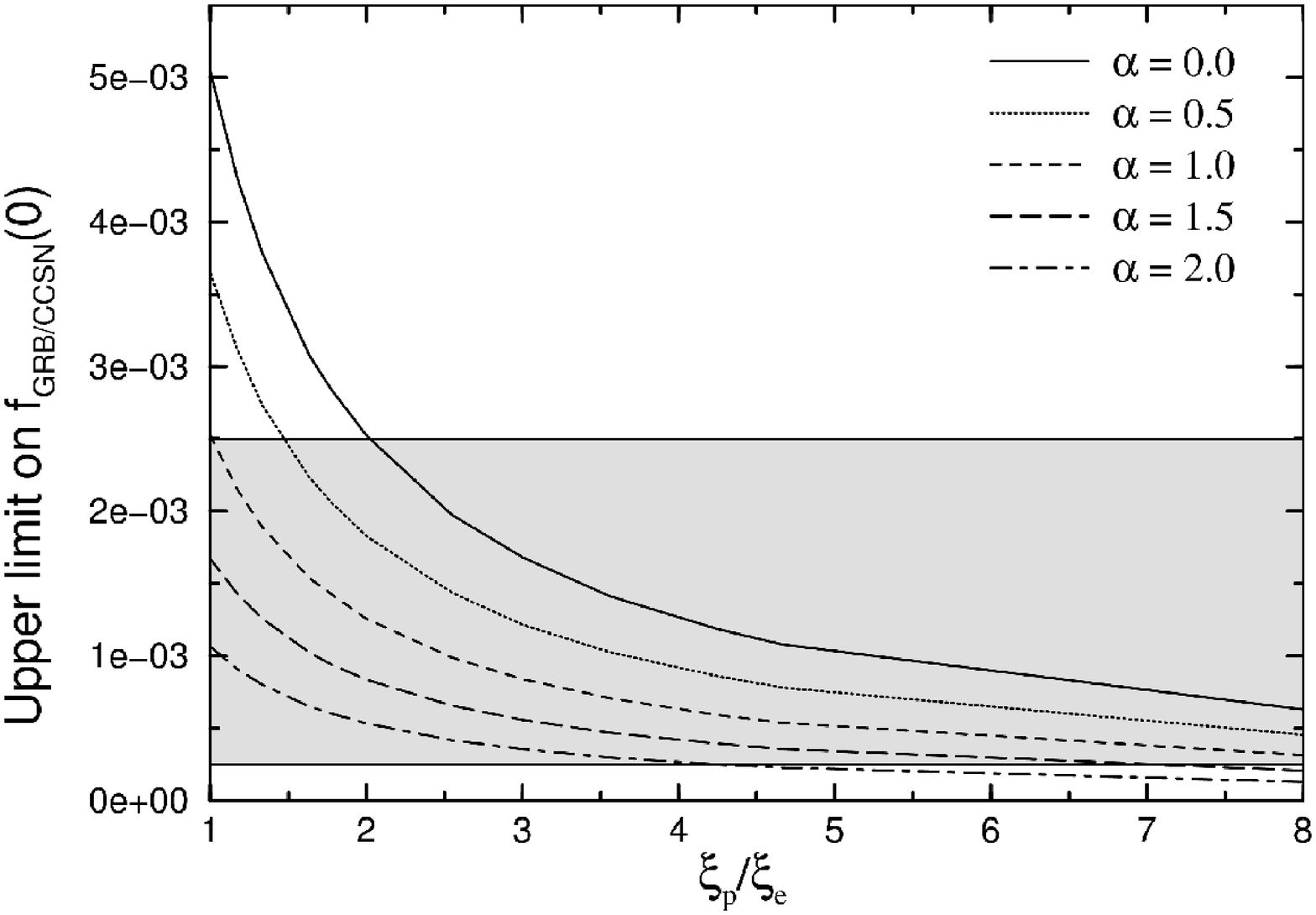,angle=270,width=4.5in}
\caption{Same as Fig.~\ref{fig:f_vs_alpha}, but as function of the 
parameter $\xi_p/\xi_e$ for different values of the GRB evolution index 
$\alpha$.} 
\label{fig:f_vs_pbye}
\end{figure}

The DGRBNuB (all flavor) flux calculated from equation 
(\ref{flux_eqn_1}) together with equations (\ref{nu_spect_1}) -- 
(\ref{enustar_eqn}), 
(\ref{rgrb_rccsn_rel}) and (\ref{rccsn_Hopkins-Beacom}) with $\fgrbccsn 
(0)=1$ and $z_{\rm max}=6$ (there is negligible contribution from $z$ 
beyond this value), and averaged over the GRB parameters in the manner 
described in the previous section, is shown 
in Figures 
\ref{fig:flux_equipart0.33} and \ref{fig:flux_alpha1.5} (the superscript 
``ob'' has been dropped in these Figures). Figure 
\ref{fig:flux_equipart0.33} shows the flux for the equipartition case of 
$\xi_p/\xi_e=1$ (i.e., $\xi_p=\xi_e=\xi_B=1/3$ with our choice of 
$\xi_e=\xi_B$) for five different values of the GRB rate  
evolution index $\alpha$ including the case $\alpha=0$ (no evolution), 
while Figure \ref{fig:flux_alpha1.5} shows the flux for 
different values of the parameter $\xi_p/\xi_e$ with $\alpha=1.5$, its  
``best-fit'' value from Ref.~\cite{Kistler_etal_07}. In both Figures, we 
also show the current all flavor 90\% C.~L.~upper limit on 
$\enu^2\Phi(\enu)$ from 
the AMANDA-II experiment~\cite{amanda_II_limit}~\cite{fn4}
and also the projected upper limit from the IceCube 
experiment after three years of operation~\cite{icecube_3year}, both for 
an assumed spectrum of the form $\Phi(\enu)\propto\enu^{-2}$.  
The resulting upper limits on $\fgrbccsn (0)$ obtained by requiring that 
$\enu^2\Phi(\enu)$ not exceed the AMANDA-II limit are shown in Figures   
\ref{fig:f_vs_alpha} and \ref{fig:f_vs_pbye} as functions of the 
parameters $\alpha$  and $\xi_p/\xi_e$, respectively. For comparison, the 
range of current estimates of the value of the ratio $\fgrbccsn (0)$ 
derived from various astronomical 
observations~\cite{Guetta-DellaValle_06,Bissaldi_etal_07} 
is also indicated in Figures \ref{fig:f_vs_alpha} and 
\ref{fig:f_vs_pbye}.  
    
It is clear that for a given value of $\alpha$ a higher value of the ratio 
$\xi_p/\xi_e$ implies a higher predicted level of DGRBNuB flux (see 
equation (\ref{nu_spect_1})), thus 
giving more stringent constraint on (i.e., a smaller upper-limit value of)   
$\fgrbccsn (0)$. Similarly, for a given value of $\xi_p/\xi_e$, a higher 
value of $\alpha$ implies more GRBs at higher redshifts, again implying a 
higher predicted level of DGRBNuB and consequently more stringent 
constraint on $\fgrbccsn (0)$. Thus, the most conservative 
limit on 
$\fgrbccsn (0)$ comes from the case $\alpha=0$ and $\xi_p/\xi_e=1$.  
These limits are $\fgrbccsn (0) \leq 5.0\times 10^{-3}$ for $\alpha=0$, 
and $\fgrbccsn (0) \leq 1.1\times 10^{-3}$ for $\alpha=2$.  
For the ``best-fit'' value of $\alpha=1.5$~\cite{Kistler_etal_07}, we have 
$\fgrbccsn (0) \leq 1.7\times 10^{-3}$ for $\xi_p/\xi_e=1$. 
We also see from Figures \ref{fig:f_vs_alpha} and 
\ref{fig:f_vs_pbye} that, for a wide range of other 
values of the parameters $\xi_p/\xi_e$ and $\alpha$,  
the upper limits on $\fgrbccsn(0)$ derived here from the consideration of 
high energy diffuse neutrino background are already 
more restrictive than the current upper limit 
($\sim 2.5\times10^{-3}$) on $\fgrbccsn(0)$ inferred from other astronomical 
considerations~\cite{Guetta-DellaValle_06,Bissaldi_etal_07}.  

At this point it should be mentioned that the AMANDA-II limit we have 
used 
above actually applies specifically to an assumed diffuse neutrino 
spectrum of the form $\Phi(\enu)\propto\enu^{-2}$. The DGRBNuB spectra 
shown in Figures \ref{fig:flux_equipart0.33} and \ref{fig:flux_alpha1.5} 
are clearly not of this form. Strictly speaking, therefore, we should 
calculate the experimental ``AMANDA-II'' upper limit for our form 
of the DGRBNuB 
spectrum and then use that to derive the upper limits on $\fgrbccsn (0)$.  
This can in principle be done by feeding the DGRBNuB 
spectra calculated above to the detailed detector simulation 
and optimized signal event selection procedures for the AMANDA 
experiment. Clearly, this is beyond our scope in this 
paper. However, use of the $\enu^{-2}$ AMANDA-II limit in 
our case here may not be too bad an approximation as a first step since, 
according to the signal event selection criteria of the 
AMANDA-II experiment as explained 
in Ref.~\cite{amanda_II_limit}, it seems reasonable to expect that the 
dominant contribution to the would-be signal events for our spectrum 
would come from the region around the broad peak 
of the $\enu^2\Phi$ spectrum where indeed $\Phi\propto \enu^{-2}$, 
approximately. Thus, while we recognize that the upper limits on 
$\fgrbccsn(0)$ derived here from directly using the $\enu^{-2}$ AMANDA-II 
limit in our case should be treated with caution, we do not expect 
significant changes in our results (by say more than a factor of few) 
under a more proper evaluation of the experimental ``AMANDA-limit'' for 
our spectrum. 

It is interesting to note from Figures \ref{fig:f_vs_alpha} and 
\ref{fig:f_vs_pbye} that the conservative upper limit on 
$\fgrbccsn(0)$ (obtained with $\xi_p/\xi_e=1$) for the case of 
$\alpha=1.5$, the best-fit value of the evolution 
parameter~\cite{Kistler_etal_07}, is not far above the current estimate of 
the lower limit on this ratio inferred from other 
considerations. For larger values of $\xi_p/\xi_e$ the upper limits are 
even closer to the otherwise estimated lower limit on 
$\fgrbccsn(0)$. This implies that the predicted
DGRBNuB flux should be detectable by the upcoming detectors such 
as IceCube 
which will have significantly improved sensitivity over that of 
AMANDA, unless the estimates of $\fgrbccsn(0)$ from direct 
astronomical observations are gross overestimates
(which is possible, for example, due to incorrect estimates of the average 
GRB beaming factor), or that the
assumption of proton acceleration to ultrahigh energies within GRB 
jets is invalid, or both of these. 

A caveat in the analysis presented above is that it is based on the 
standard assumption of variability timescales of GRBs on the 
order of milliseconds, which 
implies small emission regions and consequently large internal 
target photon densities for efficient neutrino production through 
photohadronic processes~\cite{fn5}. While millisecond timescale 
variability has been seen for many GRBs, this may not always be the 
case. Efficiency of high energy neutrino production in GRBs in the collapsar model with variability on larger timescales has been 
studied, for example, 
in Refs.~\cite{Dermer-Atoyan_03_06,Murase_etal_2006}. Also,  neutrino 
production can be effectively quenched in individual GRBs 
if $\Gamma$, the 
bulk flow Lorentz factor, is sufficiently large. Clearly, 
more precise 
determination of the distribution of the bulk flow Lorentz factor 
and variability timescale of the GRBs will be useful in 
calculating the expected level of the diffuse neutrino flux from 
GRBs more reliably which, together with the results from 
experiments such as IceCube, should be able to place more precise constraints on the fraction of 
all stellar collapse events that give rise to GRBs.      

\section{Summary and Conclusions
\label{sec:summary}}
In this paper we have attempted to derive upper limits on the fraction 
$\fgrbccsn$ of all stellar core-collapse events that give rise to GRBs, 
by using the current experimental upper limit on the high energy (TeV -- 
PeV) diffuse neutrino background given by the AMANDA-II experiment in the 
South Pole, under the assumption that GRBs are sources of such high energy 
neutrinos. High energy neutrinos are predicted to be produced within GRB 
jets through 
photopion production by protons and subsequent decay of the charged pions, 
provided protons are accelerated to ultrahigh energies at the 
internal shocks within GRB jets. In our calculation we have allowed for a 
possible evolution of the cosmic GRB rate relative to star formation rate. 
For a wide range of values of various parameters, 
the upper limits on $\fgrbccsn(0)$ derived here from the AMANDA-II 
results are already more restrictive than the upper limit on this
ratio inferred from other
astronomical considerations, thus providing a useful
independent probe of and constraint on the CCSN-GRB connection.
The closeness of the upper limits on $\fgrbccsn(0)$ derived here 
(in particular for the case of enhanced evolution of the GRB rate 
relative to the star formation rate at high redshifts) to the 
lower limit on this ratio inferred from various  
astronomical considerations seems to indicate that the 
predicted DGRBNuB 
flux should be detectable by the upcoming detectors such as 
IceCube which will have significantly improved sensitivity over that of 
AMANDA-II. On the other hand, non-detection of the DGRBNuB by 
the IceCube detector after three years of operation, for example, 
will give more stringent upper limits on $\fgrbccsn$, but at 
the same time will also imply that either the values of $\fgrbccsn$ inferred from direct astronomical observations have been significantly 
overestimated 
(which is possible, for example, due to incorrect estimates of the average
GRB beaming factor) or that the
assumption of proton acceleration to ultrahigh energies within GRB jets is
invalid, or both of these. However, more precise determination of the 
distribution of some of the crucial GRB parameters such as 
the bulk flow Lorentz factor and variability timescale of the 
GRBs will be needed to reliably calculate the expected 
contribution of the GRBs to the high energy diffuse neutrino background, 
and thereby to determine the upper limits on $\fgrbccsn$ more reliably. 
To conclude, then, the up-coming large 
volume neutrino telescopes hold immense  
promise of yielding significant information both on the nature of 
the fundamental physical process of particle acceleration in GRB sources  
as well as on the rate of occurrence of these events in the Universe.     

One of us (PB) wishes to thank Nayantara Gupta for helpful clarifications.


\begin{thebibliography}{99}
\bibitem{fn1} SNe of Type Ib (Ic) are thought to arise
from progenitors that have lost their hydrogen (and helium) envelop
before collapse, whereas the progenitor stars giving rise to
SNe of Type II have their hydrogen envelop still in place at the
time of collapse.
\bibitem{fn2} Actually, there may be cases of ``failed
supernovae'', i.e., the star may fail to explode even though core-collapse
has taken place. We shall ignore these subtleties here and use the term
``core-collapse supernova'' throughout this paper to denote all
core-collapse events.
\bibitem{fn4} We
have multiplied the single flavor ($\numu+\numubar$) AMANDA-II upper limit
given in Ref.~\cite{amanda_II_limit} by a factor of 3 to get the
all-flavor limit assuming the flavor flux ratio of the neutrinos arriving
at Earth to be 1:1:1.
\bibitem{fn5} Specifically, we have used, following 
Ref.~\cite{Gupta-Zhang_dgrbnub}, a log-normal
distribution of variability timescales with a mean at $\sim$ 38 ms and
a standard deviation of $\sim$ 30 ms.
\end{thebibliography}
\end{document}